\documentstyle[preprint,aps]{revtex}
\begin{document}

\draft

\title{Orbital Magnetism of 2D Chaotic Lattices}
\author{A.I. Nesvizhskii and A.Yu. Zyuzin \\
\em A.F.Ioffe Physico-Technical Institute,
\em St. Petersburg 194021, Russia}
\maketitle

\begin{abstract}
We study the orbital magnetism of 2D
lattices with chaotic motion of electrons withing a
primitive cell.
Using the temperature diagrammatic technique we
evaluate the averaged value and rms fluctuation of
magnetic response in the diffusive regime 
withing the model of non-interacting electrons.
The fluctuations of magnetic susceptibility
turn out to be 
large and at low temperature can be of the order of $\chi_{L}
(k_{F}l)^{3/2}$, where $k_{F}$ is the Fermi wavevector, $l$ is the
mean free path, and $\chi_{L}$ is the Landau susceptibility.   
In the certain region of magnetic fields the paramagnetic
contribution to the averaged
response is field independent and larger than the absolute value
of Landau response.
\end{abstract}

\vspace{1 cm}

In the recent years, the orbital magnetism of mesoscopic systems has
attracted much attention \cite{Oh91,Rav92,Alt93}. In the limit of the
classically weak magnetic field, when
the cyclotron radius is larger than the electron mean free
path, the structure of energy levels of mesoscopic sample is very sensitive
to the impurity configuration. As a result, the fluctuations of magnetic
response turn out to be larger than the disorder-averaged value,
i.e., the magnetic response of mesoscopic metallic sample has a random sign
as a function of impurity configuration. This makes the
investigation of sample-specific fluctuations to be of great
importance for the understanding of the orbital magnetism of
mesoscopic systems.
  
In the work of Oh et al. \cite{Oh91} the
most attention was paid to the extreme quantum coherence limit,
$L \ll L_{H}, L_{T}, \:L_{\phi}$.
Here $L$ is the linear size of sample, $L_{\phi}$ is the phase
coherent length, $ L_{H}= \sqrt{c \hbar /eH}$ is the magnetic length,  
$L_{T}= \sqrt{\hbar D/T}$ is the thermal length,
$D$ is diffusion coefficient and T is the temperature.
Mesoscopic fluctuations in the limit $L_{T} < L$ were
calculated by Raveh and Shapiro \cite{Rav92}. The averaged
magnetic response in different regimes was considered by
Altshuler et al. \cite{Alt93}. The authors of these works
dealt with the case of random distribution of impurities
all over the system.

In this letter, we study the orbital magnetism 
in 2D chaotic lattices, i.e., in periodic complex
systems with chaotic motion of electrons withing a primitive
cell. As example of such systems, we consider 2D 
metallic sample (linear size $L$) consists of
$N$ parts (linear size $a=L/ \sqrt{N}$) with
the same random configuration of impurities \cite{Tan93}.
We assume that $a \gg l$, so that the motion of electrons
withing a primitive cell is diffusive and can be
described by the diffusion constant $D$.
The system under consideration is schematically shown in Fig. 1. 

We evaluate the mesoscopic fluctuations of magnetic response from
the correlation function of the thermodynamic potential $\Omega$. The
latter is related to the correlation function of particles
number ${\cal N}$ by the following expression:
 
\begin{equation}
\langle \delta \Omega(H_1) \delta
\Omega(H_2) \rangle_{lat} = \int_{-\infty}^{\mu} \int_{-\infty}^{\mu} d \mu_1
d \mu_2 \langle \delta {\cal N}_{1} \delta {\cal N}_{2} \rangle_{lat}   \:,
\end{equation}
where $\delta {\cal N}_{i} \equiv \delta {\cal N}(\mu_{i} ;
H_{i})$, $\mu$ is the
chemical potential, $H$ is the magnetic field. 
The symbol $\langle \: \rangle_{lat}$ means
averaging over the different configurations of impurities which
have the long-ranged correlation of their positions.

The correlation function of particles number is related to the
correlation function of electron density $\rho ({\bf r})$ :
$\langle \delta {\cal N}_{1} \delta {\cal N}_{2} \rangle_{lat} =
\int_{S} \int_{S} d {\bf r}_1 d {\bf r}_2 \langle \delta \rho_{1} ({\bf r}_1) 
\delta \rho_{2} ({\bf r}_2) \rangle_{lat}$, where $S=L^2$
is the square of lattice. 
Scattering contribution to the electron density at some point ${\bf
r}_{1}$ is due to the sum over the all possible scattering paths
returning to ${\bf r}_{1}$.  The 
correlation function $\langle \delta \rho_{1} ({\bf r}_1) 
\delta \rho_{2} ({\bf r}_2) \rangle_{lat}$
is determined only by the contribution of the pairs of
phase-coherent electron paths (see, e.g. \cite{Ram91}).
In the case of lattice, an electron
traversing the closed path starting at the point ${\bf r}_1$ and
passing near the point 
${\bf r}_{2}+{\bf N}$ acquires the
same phase as an electron traversing the similar, but 
shifted in space by $ {\bf N}$, path, as it is shown in
Fig. 1. Here ${\bf N}$ is 2D vector with
components $ \{ a i,a j \}$, where integers $i,j = 0 ...
\sqrt{N}-1$ numerate the
primitive cells of lattice. Thus, calculating the sum
over ${\bf N}$ we obtain

\begin{equation}
\langle \delta {\cal N}_{1} \delta {\cal N}_{2} \rangle_{lat} = N \langle
\delta {\cal N}_{1} \delta {\cal N}_{2} \rangle_{\vec{0}} \:, 
\end{equation}
where $\langle \: \rangle_{\vec{0}}$ means averaging over
the paths shifted by ${\bf N} = \vec{0}$.   

The correlation function $\langle \delta {\cal N}_{1} \delta
{\cal N}_{2} \rangle_{{\vec{0}}}$ can be conveniently 
calculated by the temperature
diagrammatic technique. We use the diffusion approximation which
is valid when $T > \triangle$,
where $\triangle = (a^2 \nu)^{-1}$ is the mean
spacing between energy bands in chaotic lattice, $\nu$ being
the mean density of states \cite{Tan93,Ser93}. 
The largest contributions to  $\langle \delta {\cal N}_{1} \delta
{\cal N}_{2} \rangle_{{\vec{0}}}$ are the double-diffuson and double-Cooperon
diagrams. We consider magnetic fields applied
perpendicularly to 2D system and satisfying the condition
$l \ll L_{H}$. We also assume that $L_T < L$. For this case the
correlation function is

\begin{equation}
\langle \delta {\cal N}_1 \delta {\cal N}_2
\rangle_{\vec{0}} = \frac{S T}{\pi ^2 \hbar D}
\sum_{\omega > 0}^{1/ \tau} \omega Re \sum_{n \geq 0}
\sum_{s= \pm } \omega_s \left(
\omega + \hbar \omega_{s} ( n + \frac{1}{2} ) +i
\mu_{12} \right)^{-2} \: .
\end{equation}

Here $\omega = 2 \pi T m $, $m$ is an integer number, 
$\mu_{12}=\mu_1-\mu_2$, $\omega_{\pm}=2De(H_1 \pm
H_2)/(c \hbar) $, $c$ is the speed of light, $e$ is the electron charge,
$n$ is the Landau
level number. The upper cutoff in (3), $1/ \tau$, 
is due to the validity of the diffusion regime.
After substitution of (2) and (3) in (1) and some
transformations we arrive at the following expression for the correlation
function of thermodynamic potential in the form of
$t$-dependent integral:

\begin{equation}
\langle \delta \Omega(H_1) \delta
\Omega(H_2) \rangle_{lat} = \frac{N S T^2}{\pi \hbar D}
\int_{0}^{\infty} \frac{d t}{t^2 {\rm sinh}^2(\pi T t)} \left(
\frac{\omega_{+}t/2}{{\rm sinh} (\omega_{+}t/2)} +
\frac{\omega_{-}t/2}{{\rm sinh}(\omega_{-}t/2)} \right) \: .
\end{equation}

To obtain the correlation function of the
magnetic moment, one has to successively differentiate (4) on
$H_1$ and $H_2$. We would like to note that the integral in (4)
converges only after differentiation on $H_1$ and $H_2$. The
analytical form of the correlation function can be found in 
two limiting cases:

1) In the "high temperature" limit, $ \hbar \omega_{+} \ll T$,
we obtain 
\begin{equation}
\langle \delta M(H_1) \delta M(H_2) \rangle_{lat} =  
\frac{28}{5} \frac{\hbar D}{T a^2} (k_F l)^2 
\chi_{L}^2 H_1 H_2 \: .
\end{equation}
Here $\chi_L= - e^2 S/(12 \pi m^{*} c^2)$ is 2D Landau susceptibility,
$m^{*}$ is an effective electron mass.
The rms fluctuation of magnetic susceptibility 
can be directly obtained from (5):

\begin{equation}
\langle \delta \chi^2 \rangle_{lat}^{1/2} =\frac{2 \sqrt{7}}
{\sqrt{5}} \frac{L_T}{a} (k_F l) 
|\chi_{L}| \: .
\end{equation}
If impurity positions are random withing entire system (no
long-range correlation, i.e., $N=1$), $a$ should be substituted on $L$.
In this case, Eq. (6) coincides (up to a numerical factor)
with the result of Raveh and Shapiro
\cite{Rav92}. Here we would like to draw attention to the
following peculiarity of chaotic lattice. We remind that
Eq. (6) was obtained for the case $L_T/L < 1$. At the same
time, the quantity $L_T/a = N L_T/L$, appearing in the Eq.
(6), can be large. 
Indeed, the validity of the perturbative approach is $T >
\triangle$ and 
in the low temperature limit $T \sim \triangle$ 
we have \footnote{ We can consider the temperatures $T \sim
\triangle$ when it does not contradict to the condition
$L_T < L$ ($ T > \triangle (k_F l) /N$), i.e., for the
lattices with $N > (k_F l) \gg 1$. }
 $L_{T}/a \sim (\hbar D/ \triangle a^2)^{1/2} \sim (\hbar D
\nu)^{1/2} \sim (k_{F} l)^{1/2}$. Therefore, 
the rms fluctuation
of magnetic susceptibility in the case of lattice can be 
of the order of $(k_F l)^{3/2}
| \chi_L |$. 

2) In the "high field" limit, $ \hbar \omega_{+} \gg T$, we have

\begin{equation}
\langle \delta M(H_1) \delta M(H_2) \rangle_{lat} = 
\frac{162 \zeta(3)}{\pi^3}
\frac{c \hbar}{e a^2} (k_F l)^2 \chi_{L}^2
(|H_1+H_2|-|H_1-H_2|) \:,
\end{equation}
where $\zeta(x)$ is Riemann Zeta-Function.
Such form of the correlation function indicates that the magnetic
moment as well as the amplitude of its oscillations are growing
functions of magnetic field $H$. It is also interesting to
note that the correlation function $\langle \delta M(H_1)
\delta M(H_2) \rangle_{lat}$ depends only on $ min \{ H_1,
H_2 \}$. 

Now let us calculate the canonically averaged magnetic
response $\langle M \rangle_{lat}$.
It can be expressed as the sum of the diamagnetic Landau response
and a paramagnetic contribution:
$\langle M \rangle_{lat} = \chi_L H + \langle
M_{p} \rangle_{lat}$, 
where 
\begin{equation}
\langle M_{p} \rangle_{lat} = - \frac{1}{2 \nu S}
\frac{\partial}{\partial H} \langle \delta {\cal N}^2
\rangle_{lat} \: .
\end{equation}
For complete
discussion about the difference between canonical and
grand-canonical situation see \cite{Oh91,Alt93} and
references therein. Note that in the case of lattice, the number
of electrons in a primitive cell is fixed.
Calculating the correlation function  
$\langle \delta {\cal N}^2 \rangle_{lat}$ (Eqs. (2) and (3) at $\mu_1 =
\mu_2 = \mu$ and $H_1=H_2=H$) we finally obtain
\begin{equation}
\langle M_{p} \rangle_{lat} = \left\{ \begin{array}{ll}
\frac{1}{4 \pi} \frac{\triangle}{T} (k_F l) | \chi_L | H
\:,\: \mbox{if} \:\: \hbar \omega_{H} \ll T  \vspace{2mm} \\ 
\frac{{\rm ln} 2}{2 \pi^3} \frac{\triangle S e}{\hbar c} \:,\:
\mbox{if} \: \: \hbar \omega_{H} \gg T \: .
 \end{array} \right. 
\end{equation}
Here $\omega_{H} =4DeH/ ( c \hbar) $ is the Cooperon
cyclotron frequency.
In the weak magnetic fields, $\hbar \omega_{H} \ll T$
($L_H \gg L_T $) 
the averaged magnetic response is paramagnetic and
grows linearly with $H$. In stronger
magnetic fields, $\hbar \omega_{H} \gg T$
($L_H \ll L_T $), the paramagnetic
contribution $\langle M_p \rangle_{lat}$ is $H$ -
independent.
As one can see from (9), the paramagnetic contribution
$\langle M_p \rangle_{lat}$ is larger than the absolute value of
diamagnetic Landau response  $| \chi_{L} H |$
until magnetic fields 
$\hbar \omega_{H} \sim \triangle
(k_F l)$ ($L_H \sim a$). 
We note that in the case of small metallic particles the paramagnetic
contribution $\langle M_p \rangle$ becomes $H$-independent 
only in the fields, where it is small
compared to the Landau response \cite{Alt93}. 

In conclusion, we discuss the approximation that we used in
derivation of (3). In the calculation of diffusion propagators
we neglected a contribution from another type of
phase-coherent electron paths (fragment of such
trajectories shown in Fig. 2).
The contribution from the paths shifted by $2 {\bf N}$ is of
the order of weak localization correction to the diffusion
constant 
\begin{displaymath}
\frac{\delta D(2 {\bf N})}{D} \sim \frac{1}{k_F l} W({\bf
N}, \omega) \:,
\end{displaymath}
where $W({\bf R}, \omega)$ is the probability to diffuse on
the distance ${\bf R}$ during a time $\omega^{-1}$.
Because the number of such trajectories is of the order of
$(L_T/a)^{2}$, the total contribution $\sum_{{\bf N}}
\delta D(2 {\bf N})/ D \sim (L_T/a)^2 1/(k_F l) \sim \triangle/ T$. For $T >
\triangle$ we can neglect it. We note that this
correction depends on the magnetic field. This dependence
contains not only a monotonic part, but also a part which oscillates
with the period $\Phi_c / \Phi_0$, where $\Phi_c =
H a^2$
is the magnetic flux through a unit primitive cell, $\Phi_0 = hc/e$.
  
We are grateful to R. Serota for a helpful discussion.
This work is supported by the Russian Foundation for Fundamental Research
(Project number 96-02-16902).

\end{document}